\documentclass{ws-rv9x6}

\newcommand{\dst}{\displaystyle}
\newcommand{\tst}{\textstyle}
\newcommand{\rhm}{\rho_{m}}
\newcommand{\prm}{p_{m}}
\newcommand{\dph}{\dot{\phi}}
\newcommand{\ddrhm}{\frac{d\phantom{\rho}}{d\rhm}}
\newcommand{\pat}{\partial}
\newcommand{\rhmz}{\rho_{m0}}
\newcommand{\prmz}{p_{m0}}
\newcommand{\No}{N}

\begin{document}

\chapter{NON-SINGULAR COSMOLOGY AND GAUGE THEORIES OF GRAVITATION}
\addcontentsline{toc}{chapter}{Non-Singular Cosmology and Gauge
Theories of Gravitation}
%%%%%%%%%%%%%%%%%%%%%%%%%%%%%%%%%%%%%%%%%%%%%%%%%%%%%%%%%%%%%%
% title, author(s) and address(es) put here:                 %
%%%%%%%%%%%%%%%%%%%%%%%%%%%%%%%%%%%%%%%%%%%%%%%%%%%%%%%%%%%%%%

\markboth{A.\,V. Minkevich}{Non-Singular Cosmology \ldots}

\author{Albert V. Minkevich}
\addcontentsline{toc}{author}{Albert V. Minkevich}
\address{Department of Theoretical Physics, Belarussian State University,\\
av. F. Skoriny 4, 220050, Minsk, Belarus\\
Department of Physics and Computer Methods,\\
Warmia and Mazury University in Olsztyn,
 Poland\\
E-mail: MinkAV@bsu.by; awm@matman.uwm.edu.pl}

%%%%%%%%%%%%%%%%%%%%%%%%%%%%%%%%%%%%%%%%%%%%%%%%%%%%%%%%%%%%%%
% You may repeat \author \address as often as necessary      %
%%%%%%%%%%%%%%%%%%%%%%%%%%%%%%%%%%%%%%%%%%%%%%%%%%%%%%%%%%%%%%

\setcounter{figure}{0} \setcounter{equation}{0}
\setcounter{section}{0}

\begin{abstract}
The resolution of the problem of cosmological singularity in the framework of
gauge theories of gravitation is discussed. Generalized cosmological Friedmann
equations for homogeneous isotropic models filled by interacting scalar fields
and usual gravitating matter are deduced. It is shown that generic feature of
cosmological models of flat, open and closed type is their regular bouncing
character.
\end{abstract}

\section{Introduction}

As it is well known the problem of cosmological singularity (PCS) is one of the
most principal problems of general relativity theory (GR) and relativistic
cosmology. There were many attempts to resolve PCS in the frame of GR as well
as various generalizations of Einstein's theory of gravitation. A number of
regular cosmological solutions was obtained in the frame of metric theories of
gravitation and also other theories, in the frame of which gravitation is
described by using more general geometry than the Riemannian one
(see\cite{mc1,mc2,mc3} and Refs given herein). In connection with this note
that the resolution of PCS means not only obtaining regular cosmological
solutions, but also excluding singular solutions of cosmological equations, as
result generic feature of cosmological solutions has to be their regular
character. Moreover, gravitation theory and cosmological equations have to
satisfy the correspondence principle with Newton's theory of gravitation and GR
in the case of usual gravitating systems with sufficiently small energy
densities and weak gravitational fields excluding nonphysical solutions. The
greatest part of existent attempts to resolve the PCS does not satisfy
indicated conditions. Let's take for instance some examples. Although GR
permits to build regular closed inflationary models, the greatest part of
cosmological inflationary solutions of GR is singular\cite{mc4,mc5}. Metric
theories of gravitation based on gravitational Lagrangians including terms
quadratic in the curvature tensor lead to cosmological equations with high
derivatives, and although these theories permit to obtain regular cosmological
solutions with Friedmann and de Sitter asymptotics\cite{mc7}, however, they
possess nonphysical solutions also. At last time single regular cosmological
solutions were found in the frame of superstring theory (brane cosmology) by
using specific suppositions (see for example$^{8\text{--}11}$). To build cyclic
model of the Universe specific negative scalar field potential was introduced
in Refs\cite{mc12,mc13}.

As it was shown in a number of our
papers$^{14\text{--}24,\,1\text{--}3}$, the gauge theories of
gravitation (GTG) -- the Poincare GTG, metric-affine GTG --
possess important regularizing properties and permit to resolve
the PCS. Satisfying the correspondence principle with GR in the
case of usual gravitating systems with rather small energy
densities and pressures, the GTG lead to essentially different
consequences in the case of gravitating systems at extreme
conditions with extremely high energy densities and pressures,
namely the GTG lead to conclusion on possible existence of
limiting energy density for gravitating systems, nearby of which
gravitation has the character of repulsion, but not
attraction\cite{mc14}. The investigation of homogeneous isotropic
cosmological models in GTG shows, that by certain restrictions on
equation of state of gravitating matter at extreme conditions and
indefinite parameter of cosmological equations of GTG generic
feature of cosmological models including inflationary models is
their regular bouncing character. Such conclusions were obtained,
in particular, in our recent papers\cite{mc2,mc3} for inflationary
cosmological models filled by noninteracting scalar fields and
gravitating matter with linear equation of state.

Present paper is devoted to further study of regular cosmology in the frame of
GTG. In Section 2  cosmological equations for homogeneous isotropic models
filled by interacting scalar fields and gravitating matter with equation of
state in general form are deduced. In Section 3 general mathematical properties
of deduced equations and their solutions are analyzed. As illustration of
discussed theory in Section 4 particular solution for regular inflationary
cosmological model is given.

\section{Generalized cosmological Friedmann equations in GTG}

Homogeneous isotropic models in GTG are described by the following generalized
cosmological Friedmann equations (GCFE)
\begin{eqnarray}
&\displaystyle{\frac{k}{R^2}+\left\{\frac{d}{dt}\ln\left[R\sqrt{\left|1-\beta\left(\rho-
3p\right)\right|}\,\right]\right\}^2=\frac{8\pi}{3M_p^2}\,\frac{\rho-
\frac{\beta}{4}\left(\rho-3p\right)^2}{1-\beta\left(\rho-3p\right)}
\, ,}\\
&\displaystyle{\frac{\left[\dot{R}+R\left(\ln\sqrt{\left|1-\beta\left(\rho-
3p\right)\right|}\,\right)^{\cdot}\right]^\cdot}{R}=
-\frac{4\pi}{3M_p^2}\,\frac{\rho+3p+\frac{\beta}{2}\left(\rho-3p\right)^2}{
1-\beta\left(\rho-3p\right)}\, ,}
\end{eqnarray}
where $R(t)$ is the scale factor of Robertson-Walker metrics,
$k=+1,0,-1$ for closed, flat, open models respectively, $\beta$ is
indefinite parameter with inverse dimension of energy density,
$M_p$  is Planckian mass, a dot denotes differentiation with
respect to time\footnote{Parameter $\beta$ is defined as
$\beta=-\frac{1}{3}\left(16\pi\right)^2f\,M_p^{-4}$, where $f$ is
linear combination of coefficients at terms of gravitational
Lagrangian quadratic in the curvature tensor.}. (The system of
units with $\hbar=c=1$ is used). At first the GCFE were deduced in
Poincare GTG\cite{mc14}, and later it was shown that Eqs.(1)--(2)
take place also in metric-affine GTG\cite{mc25,mc26}. In the frame
of considered theory the conservation law in usual form
takes~place
\begin{equation}
\dot{\rho}+3H\left(\rho+p\right)=0,
\end{equation}
where $H=\frac{\dot{R}}{R}$   is the Hubble parameter. Besides cosmological
equations (1)--(2) gravitational equations of GTG lead to the following relation for
torsion function $S$ and nonmetricity function $Q$
\begin{equation}
S-\frac{1}{4}Q=-\frac{1}{4}\,\frac{d}{dt} \ln\left|1-\beta(\rho-3p)\right|.
\end{equation}
In Poincare GTG $Q=0$ and Eq. (4) determines the torsion function.
In metric-affine GTG there are three kinds of models\cite{mc26}:
in the Riemann-Cartan space-time ($Q=0$), in the Weyl space-time
($S=0$), in the Weyl-Cartan space-time ($S\neq 0$, $Q\neq 0$, the
function $S$ is proportional to the function $Q$). The value of
$|\beta|^{-1}$ determines the scale of extremely high energy
densities. The GCFE (1)--(2) coincide practically with Friedmann
cosmological equations of GR if the energy density is small
$\left|\beta(\rho-3p)\right|\ll 1$ ($p\neq\frac{1}{3}\rho$). The
difference between GR and GTG can be essential at extremely high
energy densities $\left|\beta(\rho-3p)\right|\gtrsim 1$.
Ultrarelativistic matter ($p=\frac{1}{3}\rho$) and gravitating
vacuum ($p=-\rho$) with constant energy density are exceptional
systems because Eqs. (1)--(2) are identical to Friedmann
cosmological equations of GR in these cases independently on
values of energy density and $S=Q=0$. The behaviour of solutions
of Eqs. (1)--(2) depends essentially on equation of state of
gravitating matter at extreme conditions and on sign of parameter
$\beta$. In the case of models filled by gravitating matter
without scalar fields the GCFE (1)--(2) lead to regular in metrics
cosmological solutions by the following restrictions: 1) $\beta>0$
and at extreme conditions $p<\frac{1}{3}\rho$, 2) $\beta<0$ and
$p>\frac{1}{3}\rho$\cite{mc1}. The investigation of models
including scalar fields on the base of Eqs (1)--(2) shows that the
choice $\beta<0$ permits to exclude the divergence of time
derivative of scalar fields\cite{mc2}. In connection with this we
put below that parameter $\beta$ is negative and in the frame of
our classical description $|\beta|^{-1} < 1\cdot M_p^4$.

By using GCFE (1)--(2) we will study homogeneous isotropic models
filled by interacting scalar field $\phi$ minimally coupled with
gravitation and gravitating matter with equation of state in
general form $p_m=p_m(\rho_m)$\footnote{The generalization for the
case with several scalar fields can be made directly.}. Then the
energy density $\rho$ and pressure $p$ take the form
\begin{equation}
\rho=\frac{1}{2}\dot{\phi}^2+V+\rho_m, \qquad p=\frac{1}{2}\dot{\phi}^2-V+p_m,
\end{equation}
where scalar field potential $V=V(\phi, \rho_m)$ includes the interaction
between scalar field and gravitating matter. In the most important particular
case of radiation (ultrarelativistic matter) the expressions of $V(\phi,
\rho_m)$ can be obtained by taking into account temperature corrections for
given scalar field potentials\cite{mc27} and the following relation for energy
density $\rhm\sim T^4$ ($T$ is absolute temperature). By using the scalar field
equation in homogeneous isotropic space
\begin{equation}
\ddot{\phi}+3H\dph=-\frac{\pat\,V}{\pat\,\phi}
\end{equation}
we obtain from Eqs. (3), (5), (6) the conservation law for gravitating matter
\begin{equation}
\dot{\rho}_m\left(1+\frac{\pat\,V}{\pat\,\rhm}\right)+3H\left(\rhm+\prm\right)=0.
\end{equation}
By virtue of Eqs.(5)--(7) the GCFE (1)--(2) can be transformed to the following
form
\begin{multline}
\label{e9}
 \left\{
 H\left[
   Z-3\beta%\left(
      \dph^2-\frac{3\beta}{2}\,\frac{\tst \rhm+\prm}{\tst 1+\frac{\pat V}{\pat \rhm}}\,
        \left(\ddrhm\,
           \left(%\vphantom{\frac{a}{a}}
              3\prm-4V
           \right)-1
        \right)
      %\right)
   \right]
 -3\beta\frac{\pat V}{\pat\phi}\dph\right\}^2
\\
+\frac{k}{R^2}\,Z^2 =\frac{8\pi}{3M_p^2}\,
 \left[
   \rhm+\frac{1}{2}\dph^2+V -\frac{1}{4}\beta\,
   \left(4V-\dph^2+\rhm-3\prm\right)^2
 \right]
\,Z
\end{multline}
\begin{multline}
\label{e10}%modified
\dot{H}\left\{
 Z-3\beta
 \left[
   \dph^2+\frac{1}{2}
   \frac{\tst \rhm+\prm}{\tst 1+\frac{\pat V}{\pat \rhm}}\,
        \left(
            \ddrhm\,\left(3\prm-4V\right)-1
        \right)
 \right]
\right\}Z
\\
+H^2 \left\{
 \left[
   Z+15\beta\dph^2-\frac{3\beta}{2}
   \frac{\tst \rhm+\prm}{\tst 1+\frac{\pat V}{\pat \rhm}}\,
        \left(\ddrhm\,
           \left(%\vphantom{\frac{a}{a}}
              3\prm-4V
           \right)-1
        \right)
   \right.\right.\\
   \left.\left.
   +\frac{9\beta}{2}\frac{\rhm+\prm}{
        \left(1+\frac{\pat V}{\pat\rhm}\right)^2}
   \left(
    \left(1+\frac{d\prm}{d\rhm}\right)
    \left(
        \ddrhm\left(3\prm-4V\right)-1
        -3\frac{\tst \rhm+\prm}{\tst 1+\frac{\pat V}{\pat \rhm}}\,
        \frac{\pat^2 V}{\pat\rhm^2}
    \right)
    \right.\right.\right.\\
    \left.\left.\left.
    +3\left(\rhm+\prm\right)\frac{d^2\prm}{d\rhm^2}
   \right)
 \right]Z
 -\frac{9}{2}\,\beta^2\,
 \left[
    \frac{\tst \rhm+\prm}{\tst 1+\frac{\pat V}{\pat \rhm}}\,
    \left(
        \ddrhm\,\left(3\prm-4V\right)-1
    \right)
    \right.\right.\\
    \left.\left.\vphantom{\frac{\tst \rhm+\prm}{\tst 1+\frac{\pat V}{\pat \rhm}}}
    +2\dph^2
 \right]^2
\right\}
%\\
-12\beta H\dph \left\{
    \frac{3\beta}{2}\,\frac{\pat V}{\pat\phi}
    \left[
        \frac{\tst \rhm+\prm}{\tst 1+\frac{\pat V}{\pat \rhm}}\,
        \left(
            \ddrhm\,\left(3\prm-4V\right)-1
        \right)
        +2\dph^2
    \right]
    \right.\\
    \left.
    -
    \left[
        \frac{\pat V}{\pat\phi}+\frac{9}{8}\,
        \frac{\tst \rhm+\prm}{\left(\tst 1+\frac{\pat V}{\pat \rhm}\right)^2}\,
        \frac{\pat^2 V}{\pat\phi\pat\rhm}
        \left(1+\frac{1}{3}\ddrhm\left(\prm+2V\right)\right)
    \right]Z
\right\}
\\
-3\beta \left[
    \frac{\pat^2 V}{\pat\phi^2}\dph^2-\left(\frac{\pat V}{\pat\phi}\right)^2
\right]Z -18\beta^2\left(\frac{\pat V}{\pat\phi}\right)^2\dph^2
\\
=\frac{8\pi}{3M_p^2} \left[
    V-\dph^2-\frac{1}{2}\left(\rhm+3\prm\right)
    -\frac{1}{4}\beta\left(4V-\dph^2+\rhm-3\prm\right)^2
\right]Z,
\end{multline}
 where $Z=1-\beta\left(4V-\dph^2+\rhm-3\prm\right)$. Relation (4) takes the form
\begin{equation}
S-\frac{1}{4}Q=\frac{3\beta}{2Z}\,
\left\{
    H
    \left[
        \dph^2+\frac{1}{2}\,\frac{\tst \rhm+\prm}{\tst 1+\frac{\pat V}{\pat \rhm}}\,
        \left(\ddrhm\,
           \left(3\prm-4V\right)-1
        \right)
    \right]
    +\frac{\pat V}{\pat \phi}\dph
\right\}.
\end{equation}
Unlike GR the cosmological equation (8) leads to essential restrictions on
admissible values of scalar field and gravitating matter. Imposing $\beta<0$,
we obtain from Eq. (8) in the case $k=0,+1$
\begin{equation}
Z\ge 0,\qquad \text{or}\qquad \dph^2\le4V+|\beta|^{-1}+\rhm-3\prm.
\end{equation}
The condition (11) is valid also for open models discussed below. This means
that relation (11) is fulfilled for all cosmological models independently on
their type $(k=0,+1,-1)$. Now let us introduce the 3-dimensional space $P$ with
axes $(\phi,\dot{\phi},\rho_m)$. The domain of admissible values of scalar
field $\phi$, time derivative $\dot{\phi}$ and energy density $\rho_m$ in space
$P$ determined by (11) is limited by bound $L$ defined as
\begin{equation} \label{e13}
Z=0\qquad \text{or}\qquad
\dot\phi=\pm\left(4V+|\beta|^{-1}+\rhm-3\prm\right)^{\frac{1}{2}}.
\end{equation}
From Eq. (8) the Hubble parameter on the bound $L$  is equal to
\begin{equation}
\label{e14}
H_{L}=-\frac{\dst \frac{\pat V}{\pat \phi}\, \dot\phi}{ \dst
\dph^2+\frac{1}{2}\,\frac{\tst \rhm+\prm}{\tst 1+
 \frac{\pat V}{\pat \rhm}}\,\left(
 \ddrhm\,
 \left(\vphantom{\frac{a}{a}} 3\prm-4V\right)-1\right)}.
\end{equation}
The right-hand part of Eq. (10) with the Hubble parameter determined by (13) is
equal to $\frac{1}{2}H$, this means that the torsion (nonmetricity) in this
case will be regular, if the Hubble parameter is regular.

\section{Solutions properties of GCFE }

Let us consider the most important general properties of cosmological solutions
of GCFE (8)--(9). At first note, by given initial conditions for variables
($\phi$, $\dot{\phi},\rho_m$) and value of $R$ there are two different
solutions corresponding to two values of the Hubble parameter following from
Eq.~(8):
\begin{equation}
\label{e15}
H_{\pm}=\frac{\dst 3\beta\,\frac{\pat V}{\pat \phi}\, \dot\phi \pm \sqrt{D}}{ \dst
Z-3\beta\left[\dph^2+\frac{1}{2}\,\frac{\tst \rhm+\prm}{\tst 1+
 \frac{\pat V}{\pat \rhm}}\,\left(
 \ddrhm\,
 \left(\vphantom{\frac{a}{a}} 3\prm-4V\right)-1\right)\right]},
\end{equation}
where
\begin{equation}
\label{e16}
D=\frac{8\pi}{3M_p^2}\,\left[\rhm+\frac{1}{2}\dph^2+V -\frac{1}{4}\beta\,
\left(4V-\dph^2+\rhm-3\prm\right)^2\right]\,Z-\frac{k}{R^2}Z^2.
\end{equation}
Unlike GR, the values of $H_{+}$ and $H_{-}$ in GTG are
sign-variable and, hence, both solutions corresponding to $H_{+}$
and $H_{-}$ can describe the expansion as well as the compression
in dependence on their sign. Below we will call solutions of GCFE
corresponding to $H_{+}$ and $H_{-}$ as $H_{+}$-solutions and
$H_{-}$-solutions respectively. Any cosmological solution contains
both $H_{-}$- and $H_{+}$-solution. In points of bound $L$ we have
$D=0$, $H_{+}=H_{-}$ and the Hubble parameter is determined
by~(13). The GCFE (8)--(9) are satisfied on the bound $L$,
corresponding solutions of GCFE -- $L$--solutions -- are their
particular solutions. The scalar field $\phi$ and energy density
$\rho_m$ for $L$-solutions can be found by integration of Eqs.
(6)--(7), where the Hubble parameter is defined by~(13), and by
using the constrain relation $Z=0$. $L$-solutions can be regular
as well singular in dependence on equation of state of gravitating
matter, scalar field potential $V$ and initial
conditions\cite{mc3}. Trajectories of particular solutions
situated on the bound $L$ have with $H_{\pm}$-solutions common
points, $H_{-}$--solutions reach the bound $L$ and
$H_{+}$-solutions originate from them, the surface $Z=0$
containing trajectories of particular solutions is envelope in
space $P$ for cosmological solutions. By using the following
formula obtained for $H_{\pm}$-solutions
 \begin{equation}
\dot{Z}=3\beta\,\frac{\dst -2\,\frac{\pat V}{\pat \phi}\, \dot\phi\, Z \mp
\sqrt{D}\left[\frac{\tst \rhm+\prm}{\tst 1+
 \frac{\pat V}{\pat \rhm}}\,\left(
 \ddrhm\,
 \left(\vphantom{\frac{a}{a}} 3\prm-4V\right)-1\right)+2\dph^2\right]}{ \dst
Z-3\beta\left[\dph^2+\frac{1}{2}\,\frac{\tst \rhm+\prm}{\tst 1+
 \frac{\pat V}{\pat \rhm}}\,\left(
 \ddrhm\,
 \left(\vphantom{\frac{a}{a}} 3\prm-4V\right)-1\right)\right]}
\end{equation}
it is easy to show that
\begin{equation}
\lim_{Z\to 0} \dot{H}_{\pm}=\dot{H}_L-\frac{8\pi}{9\beta M_p^2}\,\frac{\tst
\rhm+\frac{1}{2}\dph^2+V -\frac{1}{4}\beta\,
\left(4V-\dph^2+\rhm-3\prm\right)^2}{\tst \dph^2+\frac{1}{2}\,\frac{\tst
\rhm+\prm}{\tst 1+
 \frac{\pat V}{\pat \rhm}}\,\left(
 \ddrhm\,
 \left(\vphantom{\frac{a}{a}} 3\prm-4V\right)-1\right)}.
 \end{equation}
From (17) follows that in points of bound $L$ the derivatives $\dot{H}_{+}$ and
$\dot{H}_{-}$ are equal and their values do not depend on the model type, as
result we have the smooth transition from $H_{-}$-solution to $H_{+}$-solution
on bound $L$, and corresponding cosmological solutions for all types models are
regular in metrics, Hubble parameter and its time derivative. At the same time
according to (17) the value of the time derivative
 $\dot{H}_L$ for $L$-solutions is not equal to
$lim_{Z\to 0} \dot{H}_{\pm}$ and smooth transition from $H_{-}$-solutions to
$L$-solutions and from $L$-solutions to $H_{+}$-solutions without jump of the
derivative $\dot{H}$ is impossible. Note, that according to Eq. (16) the functions
$S$ and $Q$ have the following asymptotics for $H_{+}$- and $H_{-}$-solutions at
$Z\to 0$:
\begin{equation}
\lim_{Z\to 0}(S-\frac{1}{4}\,Q)=-\frac{1}{4}\,\lim_{Z\to
0}\frac{\dot{Z}}{Z}\sim\mp Z^{-\frac{1}{2}}
\end{equation}
Unlike flat and open models, for which $H_{+}= H_{-}$ only in
points of bound $L$ and regular inflationary models include
$H_{+}$-- and $H_{-}$--solutions reaching bound $L$, in the case
of closed models the regular transition from $H_{-}$-solution to
$H_{+}$-solution is possible without reaching the bound $L$. It is
because by certain value of $R$ according to (14)--(15) we have
$H_{+}= H_{-}$ in the case $Z\neq 0$. Such models are regular also
in torsion and/or nonmetricity. Regular inflationary solution of
such type was considered in Ref.\cite{mc6}.

 In order to study the behaviour
of cosmological models at the beginning of cosmological expansion, let us
analyze extreme points for the scale factor $R(t)$: $R_0=R(0)$, ${H_0=H(0)=0}$.
(This means that in the case of $H_{+}$--solutions $H_{+0}=0$ and in the case
of $H_{-}$--solutions $H_{-0}=0$). Denoting values of quantities at $t=0$ by
means of index "0", we obtain from (8)--(9):
\begin{multline}
\label{e17}
\frac{k}{R_0^2}\,Z_0^2 +9\beta^2{\left(\frac{\pat V}{\pat\phi}\right)\!}_0^2\dph_0^2
\\
=\frac{8\pi}{3M_p^2}\,
 \left[
   \rhmz+\frac{1}{2}\dph_0^2+V_0 -\frac{1}{4}\beta\,
   \left(4V_0-\dph_0^2+\rhmz-3\prmz\right)^2
 \right]
\,Z_0,
\end{multline}
\begin{multline}
\label{e18}
\dot{H}_0=
\left\{
    \frac{8\pi}{3M_p^2}
    \left[
        V_0-\dph_0^2-\frac{1}{2}\left(\rhmz+3\prmz\right)
        \right.\right.\\
        \left.\left.
        -\frac{1}{4}\beta\left(4V_0-\dph_0^2+\rhmz-3\prmz\right)^2
    \right]
    \right.\\
    \left.
    +3\beta
    \left[
        {\left(\frac{\pat^2 V}{\pat\phi^2}\right)\!}_0\dph_0^2
        -{\left(\frac{\pat V}{\pat\phi}\right)\!}_0^2
    \right]
    +18\beta^2{\left(\frac{\pat V}{\pat\phi}\right)\!}_0^2\dph_0^2\:Z_0^{-1}
\right\}
\\
\times
\left\{
 Z_0-3\beta
 \left[
   \dph_0^2+\frac{1}{2}\,\frac{\tst \rhmz+\prmz}{\tst 1+\left(\frac{\pat V}{\pat \rhm}\right)_0}\,
        \left(
            \ddrhm\,\left(3\prm-4V\right)_0-1
        \right)
 \right]
\right\}^{-1},
\end{multline}
where $Z_0=1-\beta\left(4V_0-\dot{\phi}_0^2+\rho_{m0}-3p_{m0}\right)$. A bounce
point is described by Eq. (19), if the value of $\dot{H}_0$ is positive. By using
Eq.(19) we can rewrite the expression of $\dot{H}_0$ in the form
\begin{eqnarray}
&\displaystyle{  \dot{H}_0= \left\{
    \frac{8\pi}{M_p^2}
    \left[
        V_0+\frac{1}{2}\left(\rhmz-\prmz\right)
        %\right.\right.\\
        %\left.\left.
        -\frac{1}{4}\beta\left(4V_0-\dph_0^2+\rhmz-3\prmz\right)^2
    \right]\nonumber
    \right.}\\
    &\displaystyle{
    \left.
    +3\beta
    \left[
        {\left(\frac{\pat^2 V}{\pat\phi^2}\right)\!}_0\dph_0^2
        -{\left(\frac{\pat V}{\pat\phi}\right)\!}_0^2
    \right]
    -\frac{2k}{R_0^2}\,Z_0^2
\right\}}\nonumber
\\&\displaystyle{
\times \left\{
 Z_0-3\beta
 \left[
   \dph_0^2+\frac{1}{2}\,\frac{\tst \rhmz+\prmz}{\tst 1+\left(\frac{\pat V}{\pat \rhm}\right)_0}\,
        \left(
            \ddrhm\,\left(3\prm-4V\right)_0-1
        \right)
 \right]
\right\}^{-1}\!.}\label{e18a}
\end{eqnarray}

We see from (21) unlike GR the presence of gravitating matter
(with $p_m\le \rho_m$) does not prevent from the bounce
realization\footnote{In GR a bounce is possible only in closed
models if the following condition
$V_0-\dot{\phi}_0^2-\frac{1}{2}(\rho_{m0}+3p_{mo})>0$ takes
place.} Eq. (19) determines in space $P$ extremum surfaces
depending on the value of $\beta$ and in the case of closed and
open models also parametrically on the scale factor $R_0$. In the
case of various scalar field potentials applying in inflationary
cosmology the value of $\dot{H}_{+0}$ or $\dot{H}_{-0}$ is
positive on the greatest part of extremum surfaces, which can be
called "bounce surfaces". \footnote{If $\left|\beta\right|^{-1}\ll
M_p^4$ the derivative $\dot{H}_0$ is negative in the neighbourhood
of origin of coordinates in space $P$ that leads to appearance of
oscillating solutions of GCFE\cite{mc28}.}. By giving concrete
form of potential $V$ and choosing values of $R_0$, $\phi_0$,
$\dot{\phi}_0$ and $\rho_{m0}$ at a bounce, we can obtain
numerically particular bouncing solutions of GCFE for various
values of parameter $\beta$.

The analysis of GCFE shows, that properties of cosmological solutions depend
essentially on parameter  $\beta$, i.e. on the scale of extremely high energy
densities. From physical point of view interesting results can be obtained, if
the value of $|\beta|^{-1}$   is much less than the Planckian energy density
\cite{mc6}, i.e. in the case of large in module values of parameter $\beta$ (by
imposing $M_p=1$). In order to investigate cosmological solutions at the
beginning of cosmological expansion in this case, let us consider the GCFE by
supposing that
\begin{eqnarray}
\label{e20}
&\displaystyle{\left|\beta\left(4V-\dph^2+\rhm-3\prm\right)\right|\gg
1,\nonumber}
\\
&\displaystyle{\rhm+\frac{1}{2}\dph^2+V\ll \left|\beta\right|
\left(4V-\dph^2+\rhm-3\prm\right)^2.}
\end{eqnarray}
Note that the second condition (22) does not exclude that
ultrarelativistic matter energy density can dominate at a bounce.
We obtain: {\allowdisplaybreaks
\begin{multline}
\label{e21}%modified
\frac{k}{R^2}\left(4V-\dph^2+\rhm-3\prm\right)^2+
\\
\left\{H
    \left[
        4V+2\dph^2+\rhm-3\prm
        +\frac{3}{2}\,\frac{\tst \rhm+\prm}{\tst 1+\frac{\pat V}{\pat \rhm}}\,
        \left(
            \ddrhm\,\left(3\prm-4V\right)-1
        \right)
    \right]
    \right.\\
    \left.\vphantom{\frac{\tst \rhm+\prm}{\tst 1+\frac{\pat V}{\pat \rhm}}}
    +3\frac{\pat V}{\pat\phi}\dph
\right\}^2 =\frac{2\pi}{3M_p^2}\,\left(4V-\dph^2+\rhm-3\prm\right)^3
\end{multline}
\begin{multline}
\label{e22} %modified
\dot{H} \left[
    4V+2\dph^2+\rhm-3\prm
    -\frac{3}{2}\frac{\tst \rhm+\prm}{\tst 1+\frac{\pat V}{\pat \rhm}}\,
    \left(
        \ddrhm\,\left(3\prm-4V\right)-1
    \right)
\right]
\\
\times\left(4V-\dph^2+\rhm-3\prm\right) +H^2 \left\{
    \left[
        4V-16\dph^2+\rhm-3\prm
        \vphantom{\frac{\tst \rhm+\prm}{\tst 1+\frac{\pat V}{\pat \rhm}}}
        \right.\right.\\
        \left.\left.
        +\frac{3}{2}
        \frac{\tst \rhm+\prm}{\tst 1+\frac{\pat V}{\pat \rhm}}\,
        \left(
            \ddrhm\,\left(3\prm-4V\right)-1
        \right)
        -\frac{9}{2}\frac{\rhm+\prm}{
            \left(1+\frac{\pat V}{\pat\rhm}\right)^2}
        \right.\right.\\
        \left.\left.\times
        \left(
            \left(1+\frac{d\prm}{d\rhm}\right)
            \left(
                \ddrhm\left(3\prm-4V\right)-1
                -3\frac{\tst \rhm+\prm}{\tst 1+\frac{\pat V}{\pat \rhm}}\,
                \frac{\pat^2 V}{\pat\rhm^2}
            \right)
            \right.\right.\right.\\
            \left.\left.\left.
            +3\left(\rhm+\prm\right)\frac{d^2\prm}{d\rhm^2}
        \right)
    \right]
    \left(4V-\dph^2+\rhm-3\prm\right)
    \right.\\
    \left.
    -\frac{9}{2}\,
    \left[
        \frac{\tst \rhm+\prm}{\tst 1+\frac{\pat V}{\pat \rhm}}\,
        \left(
            \ddrhm\,\left(3\prm-4V\right)-1
        \right)+2\dph^2
    \right]^2
\right\}
\\
-12H\dph \left\{
    \left[
        \frac{\pat V}{\pat\phi}+\frac{9}{8}\,
        \frac{\tst \rhm+\prm}{\left(\tst 1+\frac{\pat V}{\pat \rhm}\right)^2}\,
        \frac{\pat^2 V}{\pat\phi\pat\rhm}
        \left(1+\frac{1}{3}\ddrhm\left(\prm+2V\right)\right)
    \right]
    \right.\\
    \left.\times
    \left(4V-\dph^2+\rhm-3\prm\right)
    +\frac{3}{2}\,\frac{\pat V}{\pat\phi}
    \left[
        \frac{\tst \rhm+\prm}{\tst 1+\frac{\pat V}{\pat \rhm}}\,
        \left(
            \ddrhm\,\left(3\prm-4V\right)-1
        \right)
        \right.\right.\\
        \left.\left.
        \vphantom{\frac{\tst \rhm+\prm}{\tst 1+\frac{\pat V}{\pat \rhm}}}
        +2\dph^2
    \right]
\right\} +3 \left[
    \frac{\pat^2 V}{\pat\phi^2}\dph^2-\left(\frac{\pat V}{\pat\phi}\right)^2
\right]\left(4V-\dph^2+\rhm-3\prm\right)
\\
-18\left(\frac{\pat V}{\pat\phi}\right)^2\dph^2 =\frac{2\pi}{3M_p^2} \left(
    4V-\dph^2+\rhm-3\prm
\right)^3,
\end{multline}
}
 Equations (23)--(24) do not contain the parameter $\beta$.
According to Eq. (23) the Hubble parameter in considered
approximation is equal to
\begin{multline}
\label{e23}%modified
H_{\pm}= \left[
    4V+2\dph^2+\rhm-3\prm
    +\frac{3}{2}\,\frac{\tst \rhm+\prm}{\tst 1+\frac{\pat V}{\pat \rhm}}\,
    \left(
        \ddrhm\,\left(3\prm-4V\right)-1
    \right)
\right]^{-1}
\\
\!\times \left[
    -3\frac{\pat V}{\pat \phi}\, \dph\pm
    \left|4V-\dph^2+\rhm-3\prm\right|
    \sqrt{
        \frac{2\pi}{3M_p^2}\left(4V-\dph^2+\rhm-3\prm\right)
        -\frac{k}{R^2}
    }\,\,
\right]
\\{}\!
\end{multline}
and  extreme points of the scale factor are determined by the following
condition
\begin{equation}
\label{e24}
\frac{k}{R_0^2}+9
\left[
    \frac{{\left(\dst \frac{\pat V}{\pat\phi}\right)\!}_0\dph_0}
        {\dst 4V_0-\dph_0^2+\rhmz-3\prmz}
\right]^2 =\frac{2\pi}{3M_p^2}\,\left(4V_0-\dph_0^2+\rhmz-3\prmz\right).
\end{equation}
From Eq. (24) the time derivative of the Hubble parameter at
extreme points is
 {\allowdisplaybreaks
\begin{multline}
\label{e25}
\dot{H}_0 =
\left\{
    \frac{2\pi}{3M_p^2} \left(4V_0-\dph_0^2+\rhmz-3\prmz\right)^2
    \right.\\
    \left.
    -3
    \left[
       {\left(\frac{\pat^2 V}{\pat\phi^2}\right)\!}_0\dph_0^2
       -{\left(
            \frac{\pat V}{\pat\phi}
        \right)\!}_0^2%
    \right]
    +\frac{18{\left(\frac{\pat V}{\pat\phi}\right)\!}_0^2\dph^2}{
        4V_0-\dph_0^2+\rhmz-3\prmz}
\right\}
\\
\times
\left[
    4V_0+2\dph_0^2+\rhmz-3\prmz
    -\frac{3}{2}\frac{\tst \rhmz+\prmz}{\tst 1+{\left(\frac{\pat V}{\pat \rhm}\right)\!}_0}\,
    \left(
        \ddrhm\,\left({3\prm-4V\!}\right)_0-1
    \right)
\right]^{-1}
\end{multline}
 }
Obviously Eqs. (26)--(27) correspond to (19)--(20) in considered approximation.

The analysis given in this Section confirms conclusions obtained
in our previous papers\cite{mc2,mc3}. The existence of limiting
bound $L$ and bounce surface in space $P$ in the case of models
including scalar fields and gravitating matter ensures regular
character of cosmological solutions in metrics, Hubble parameter
and its time derivative; corresponding restriction on equation of
state of gravitating matter and potential $V$ at extreme
conditions is: $\dst \left(1+\frac{\pat V}{\pat
\rhm}\right)^{-1}\left(\ddrhm\,\left(3\prm-4V\right)-1\right)>0$
(see (14)). In absence of scalar fields the bound $L$ is reduced
to a point, particular solution of GCFE is stationary ($H$=0), and
the condition $Z=0$ determines a bounce point and the value of
limiting energy density simultaneously if at extreme conditions
$\prm>\frac{1}{3}\rhm$\cite{mc1}.

\section{Regular inflationary cosmological models}

\begin{figure}[htb!]
\begin{minipage}{0.48\textwidth}\centering{
\epsfig{file=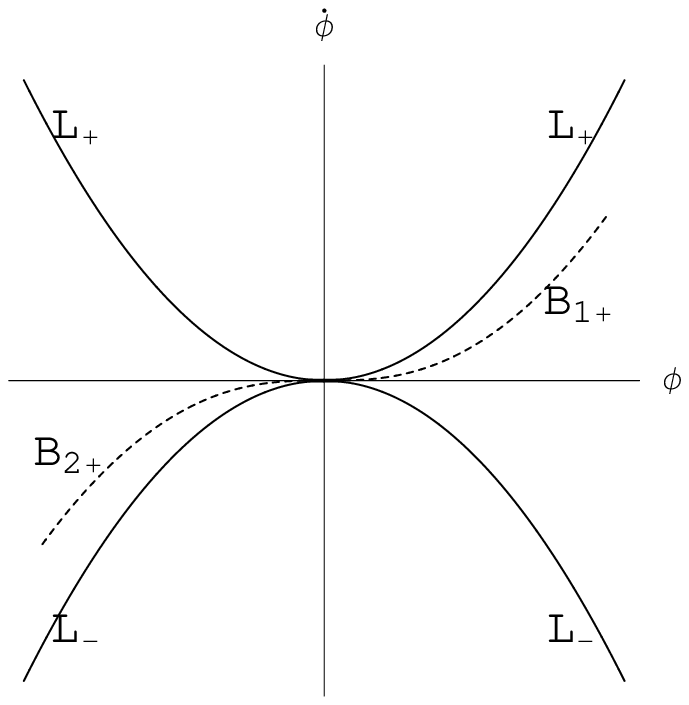,width=\linewidth}}
\end{minipage}\, \hfill\,
\begin{minipage}{0.48\textwidth}\centering{
\epsfig{file=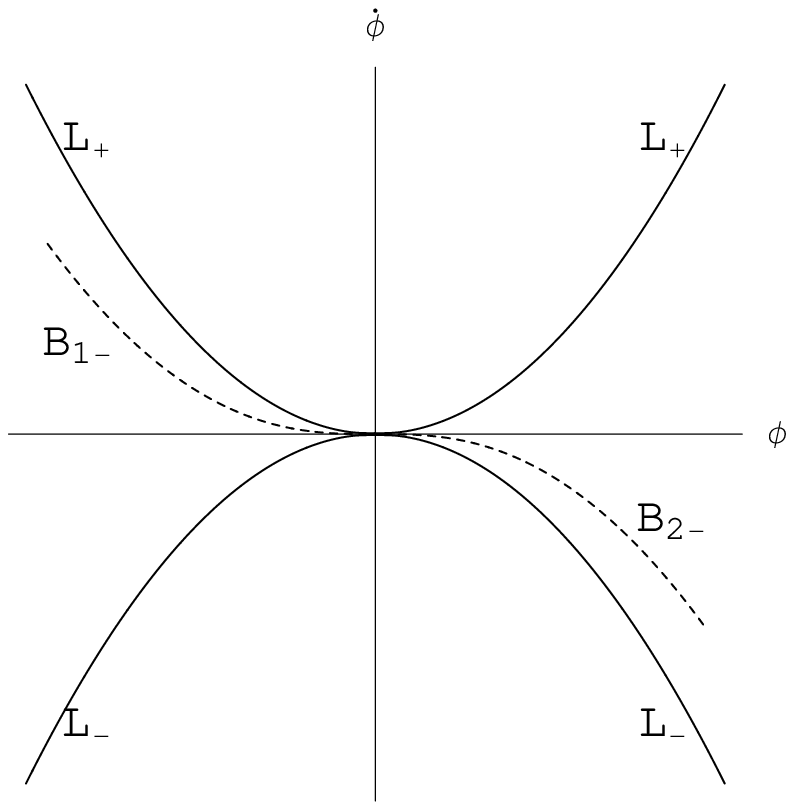,width=\linewidth}}
\end{minipage}
\caption[]{\label{figm1} Bound $L_\pm$-curves and bounce curves for flat models in
the case of potential $V=\frac{1}{4}\lambda\phi^4$.}
\end{figure}

As illustration of discussed theory regular cosmological models in the simplest
particular case will be considered in this Section. We will consider models
including noninteracting scalar field with potential $V(\phi)$ and
ultrarelativistic matter ($\prm=\frac{1}{3}\rhm$)\cite{mc2}. In this case the
bound $L$ in space $P$ is reduced to two cylindric surfaces
$\dot{\phi}=\pm\left(4V+|\beta|^{-1}\right)^{\frac{1}{2}}$. Bounce surface is
reduced also to cylindric surfaces in the case under consideration, when the
scale of extremely high energy densities is much smaller than the Planckian
energy density (see (26)). In connection with this we will consider instead of
space $P$ the plane of variables ($\phi$, $\dot\phi$) and intersections of the
bound $L$ and bounce surface with this plane. We have in this plane two bound
$L_\pm$--curves and in the case of flat models two bounce curves $B_1$ and
$B_2$ determined by equation\footnote{The neighbourhood of origin of
coordinates is not considered in this approximation, the behavior of bounce
curves near origin of coordinates was examined in Ref.~\cite{mc28}, where
scalar fields superdense gravitating systems were discussed.}
\[
4V_0-\dot{\phi}_0^2=3\left(\frac{M_p^2}{2\pi}\,{V_0'}^{2}\,\dot{\phi}_0^2\right)^{\frac{1}{3}}.
\]
Each of two curves $B_{1,2}$ contains two parts corresponding to vanishing of
$H_{+}$ or $H_{-}$ and denoting by ($B_{1+}$, $B_{2+}$) and ($B_{1-}$,
$B_{2-}$) respectively. If $V'$ is positive (negative) in quadrants 1 and 4 (2
and 3) on the plane ($\phi$, $\dot\phi$), the bounce will take place in points
of bounce curves $B_{1+}$ and $B_{2+}$ ($B_{1-}$ and $B_{2-}$) in quadrants 1
and 3 (2 and 4) for $H_{+}$-solutions ($H_{-}$-solutions) (see
Fig.~\ref{figm1}).
To analyze flat bouncing models we have to take into account  that besides
regions lying between curves $L_\pm$ and corresponding bounce curves the sign
of values $H_{+}$ and $H_{-}$ for applying potentials is normal: $H_{+}>0$,
$H_{-}<0$. The Hubble parameter $H_{+}$ is negative in regions between curves
($L_{+}$ and $B_{1+}$), ($L_{-}$ and $B_{2+}$), and the value of $H_{-}$ is
positive in regions between curves ($L_{+}$ and $B_{1-}$), ($L_{-}$ and
$B_{2-}$). As it was noted above any cosmological solution has to contain both
$H_{-}$- and $H_{+}$-solution. The regular transition from $H_{-}$-solution to
$H_{+}$-solution takes place in points of $L_{\pm}$ where $H_{+}=H_{-}$. In the
case of open and closed models Eq.(26) determines 1-parametric family of bounce
curves with parameter $R_0$. Bounce curves of closed models are situated
 in region between two bounce curves $B_{1}$ and $B_{2}$ of
flat models, and in the case of open models bounce curves are situated in two
regions between the curves: $L_{+}$ and $B_{1}$, $L_{-}$ and $B_{2}$. In
general case, when approximation (22) is not valid, bounce surface in space $P$
of cosmological models including scalar field and ultrarelativistic matter
determined by Eq.(19) in space $P$ depends on parameter $\beta$ and it is not
more cylindric surface. The situation concerning cosmological solutions of Eqs.
(8)--(9) does not change.

 Note that the analysis of inflationary solutions reaching the bound
$L$ by numerical integration of Eqs. (9) and (6) is difficult, because the
coefficient at $\dot{H}$ in Eq. (9) tends to zero at $Z\to 0$.

\begin{figure}[htb!]
\begin{minipage}{0.48\textwidth}\centering{
\epsfig{file=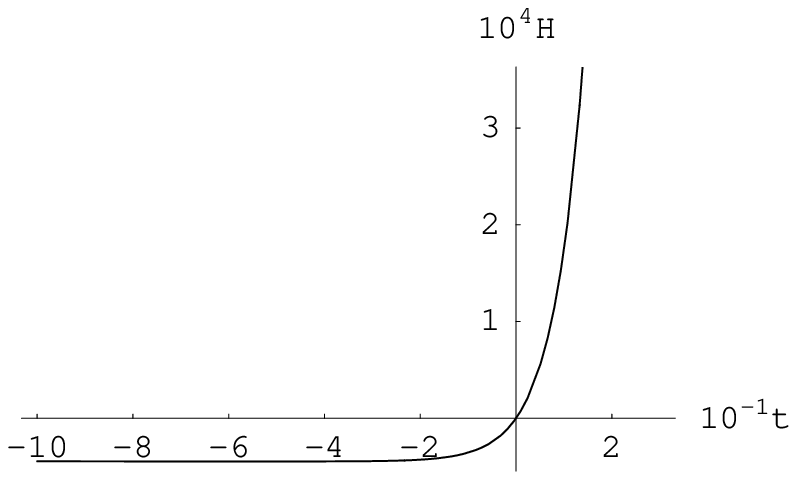,width=\linewidth}}
\end{minipage}\, \hfill\,
\begin{minipage}{0.48\textwidth}\centering{
\epsfig{file=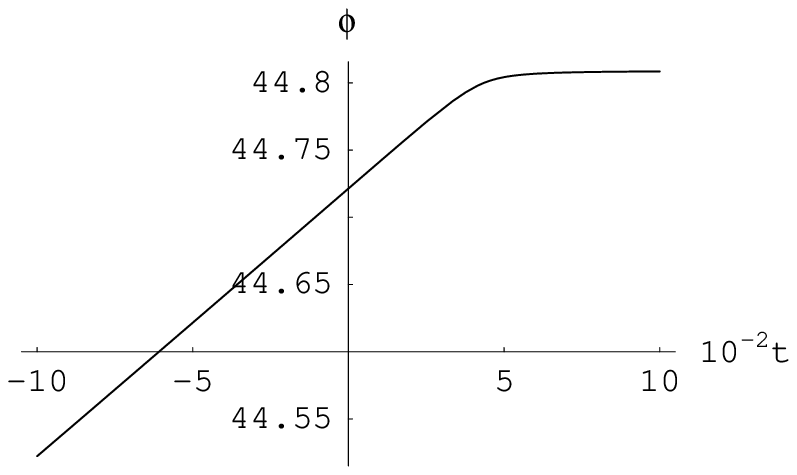,width=\linewidth}}
\end{minipage}%\\
\caption{\label{figm2}The stage of transition from compression to expansion.}
\end{figure}

\begin{figure}[htb!]
\begin{minipage}{0.48\textwidth}\centering{
\epsfig{file=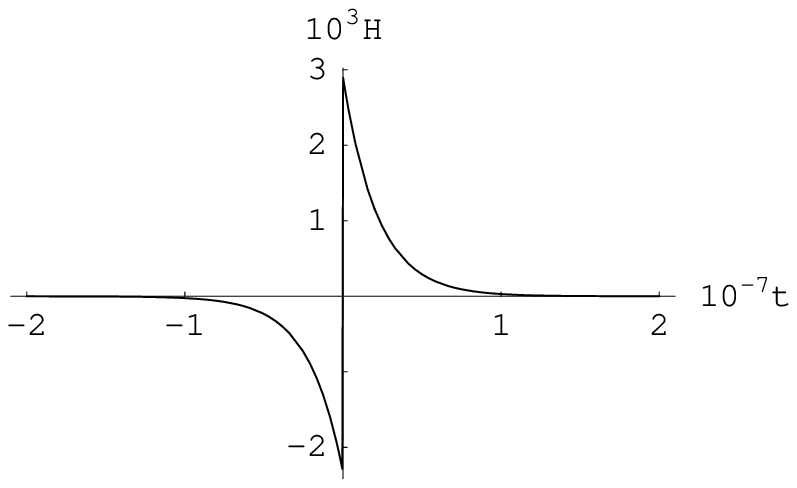,width=\linewidth}}
\end{minipage}\, \hfill\,
\begin{minipage}{0.48\textwidth}\centering{
\epsfig{file=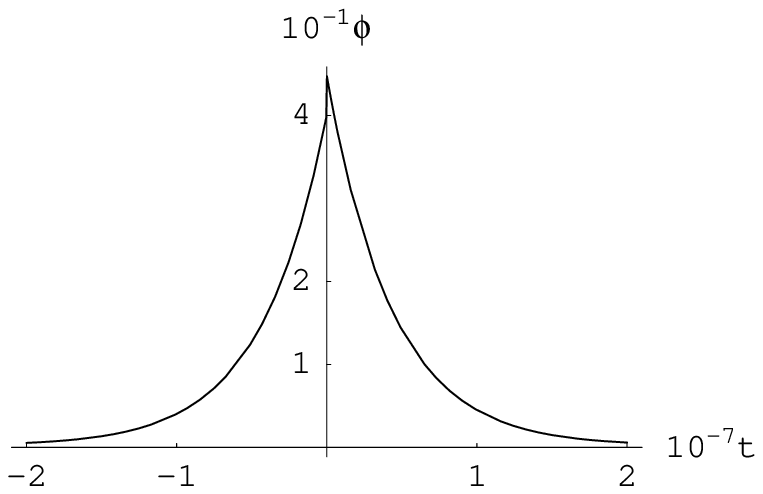,width=\linewidth}}
\end{minipage}
\caption{\label{figm3}Quasi-de-Sitter stage of compression and inflationary stage.}
\end{figure}

\begin{figure}[htb!]
\begin{minipage}{0.48\textwidth}\centering{
\epsfig{file=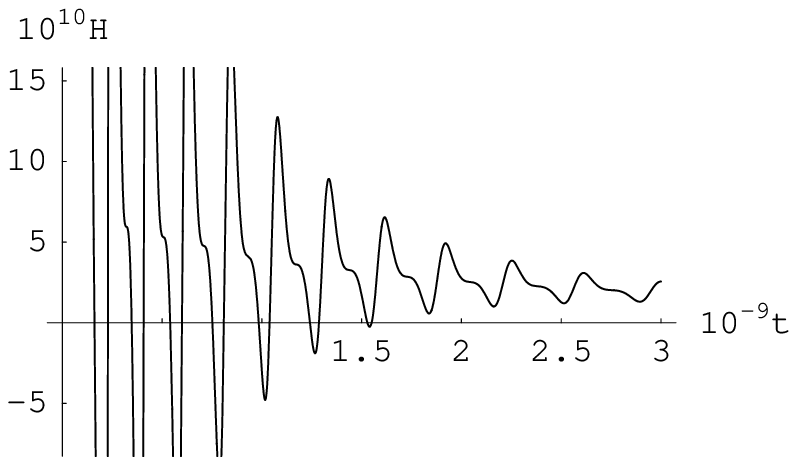,width=\linewidth}}
\end{minipage}\, \hfill\,
\begin{minipage}{0.48\textwidth}\centering{
\epsfig{file=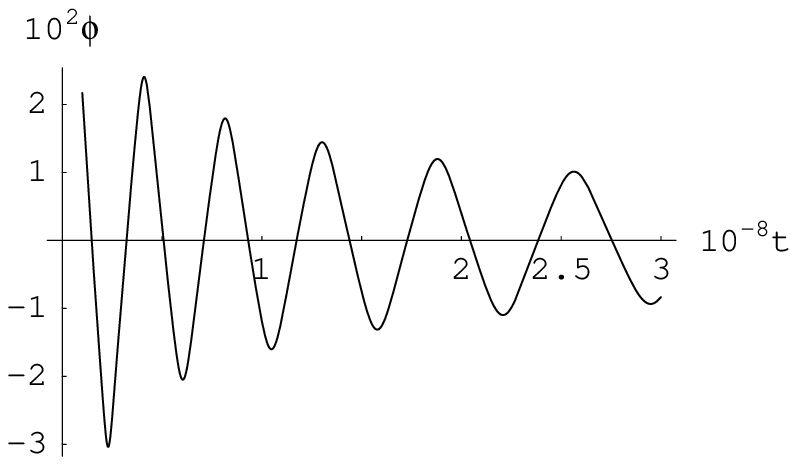,width=\linewidth}}
\end{minipage}
\caption{\label{figm4}The stage after inflation.}
\end{figure}

Below particular bouncing cosmological inflationary solution for flat model by using
scalar field potential in the form $V=\frac{1}{4}\lambda\phi^4$
(${\lambda=10^{-12}}$) is given. The solution was obtained by numerical integration
of Eqs. (6), (9) and by choosing in accordance with Eq.(19) (or (26)) the following
initial conditions at a bounce: $\phi_0=\sqrt{2}\cdot 10^{3/2}\, M_p$,
$\dot{\phi}_0=1,998\cdot 10^{-3} M_p^2$ ($\beta=-10^{18}M_p^{-4}$); initial value of
$R_0$ can be arbitrary. As was noted above, the radiation energy density does not
have influence on the dynamics of cosmological model, if the value of $\rho_m$
satisfies the condition (22). A bouncing solution includes: quasi-de-Sitter stage of
compression, the stage of transition from compression to expansion, quasi-de-Sitter
inflationary stage, stage after inflation. The dynamics of the Hubble parameter and
scalar field is presented for different stages of obtained bouncing solution in
Figures \ref{figm2}--\ref{figm4} (by choosing $M_p=1$).
The transition stage from compression to expansion
(Fig.~\ref{figm2}) is essentially asymmetric with respect to the
point $t=0$ because of $\dot{\phi}_0\neq 0$. In course of
transition stage the Hubble parameter changes from maximum in
module negative value at the end of compression stage to maximum
positive value at the beginning of expansion stage. The scalar
field changes linearly at most part of transition stage, the
derivative $\dot\phi$ grows at first being positive to maximum
value $\dot{\phi}\sim\dot{\phi}_0$ and then the value of
$\dot\phi$ decreases and becomes negative. Quasi-de-Sitter
inflationary stage and quasi-de-Sitter compression stage are
presented in Fig.~\ref{figm3}. The amplitude and frequency of
oscillating scalar field after inflation (Fig.~\ref{figm4}) are
different than that of GR, this means that approximation of small
energy densities
$\left|\beta\left(4V-\dot{\phi}^2-2\rho_1\right)\right|\ll 1$ at
the beginning of this stage is not valid; however, the
approximation (23)--(24) is not valid also because of dependence
on parameter $\beta$  of oscillations characteristics\cite{mc6}.
The behaviour of the Hubble parameter after inflation is also
noneinsteinian, at first the Hubble parameter oscillates near the
value $H=0$, and later the Hubble parameter becomes positive and
decreases with the time like in GR. Before quasi-de-Sitter
compression stage there are also oscillations of the Hubble
parameter and scalar field not presented in Figures
\ref{figm2}--\ref{figm4}. Ultrarelativistic matter, which could
dominate at a bounce has negligibly small energy densities at
quasi-de Sitter stages. At the same time the gravitating matter
could be at compression stage in more realistic bouncing models,
and scalar fields could appear only at certain stage of
cosmological compression. As it follows from our consideration
regular character of such inflationary cosmological models has to
be ensured by cosmological equations of GTG.

The interaction between scalar fields and radiation leads to quantitative
corrections of considered inflationary cosmological models. It is necessary to
note that corrections for scalar field potentials quadratic in the temperature
$T^2\sim\rhm^{1/2}$ can play the important role for more late stages of
cosmological evolution, when energy densities are sufficiently small and
consequences of GTG and GR coincide. In accordance with Eq. (7) these terms
change essentially the connection between scale factor $R$ and radiation energy
density. By certain restrictions on parameters of potentials $V$ indicated
terms can be greater than radiation energy density and scalar field energy
density (without terms of interaction) at late stages of cosmological expansion
 leading to acceleration of cosmological
expansion\cite{mc29}.

Note that bouncing character have solutions not only in classical region, where
scalar field potential, kinetic energy density of scalar field and energy
density of gravitating matter do not exceed the Planckian energy density, but
also in regions, where classical restrictions are not fulfilled and according
to accepted opinion quantum gravitational effects can be essential.
\section{Conclusion}

As it is shown in our paper, GTG  permit to build non-singular cosmology and at
first of all regular inflationary cosmology, if gravitating matter and scalar
fields satisfy certain restrictions at extreme conditions and indefinite
parameter $\beta$ is negative. All cosmological solutions for flat, open and
closed models are regular in metrics, Hubble parameter, its time derivative
and, hence, not limited in the time. The presence of scalar fields leading to
appearance of inflation in cosmological models changes essentially the
structure of GCFE, as result a family of closed models regular also in torsion
and/or nonmetricity appears. To build realistic
 cosmological models we have to know properties of matter filling the evolving Universe,
and at first of all the change of equation of state of gravitating matter and
properties of scalar fields by evolution of the Universe.

\section*{Acknowledgements}
\addcontentsline{toc}{section}{Acknowledgements}

I am grateful to  Alexander Garkun and  Andrey Minkevich for  help by
preparation of this paper. I am thankful to my wife Eleonora for her permanent
support.

%%%%%%%%%%%%%%%%%%%%%%%%%%%%%%%%%%%%%%%%%%%%%%%%%%%%%%%%%%%%%
% Doing references:                                         %
%%%%%%%%%%%%%%%%%%%%%%%%%%%%%%%%%%%%%%%%%%%%%%%%%%%%%%%%%%%%%

%\setlength{\textfloatsep}{\the\oldtextfloatsep}
%\setlength{\intextsep}{\the\oldintextsep}

%\bibitem{ja} M. Barranco and J. R. Buchler, {\it Phys. Rev.}
%{\bf C34}, 1729 (1980).

%\bibitem{ppz} G. Pang and H. Zhao, {\it J. Phys. A: Math. Gen.}
%{\bf 25}, L527 (1992).

\end{document}